\documentclass[12pt]{article}

\def\square{\kern1pt\vbox{\hrule height 1.2pt\hbox{\vrule width 1.2pt\hskip 3pt
   \vbox{\vskip 6pt}\hskip 3pt\vrule width 0.6pt}\hrule height 0.6pt}\kern1pt}

\begin{document}

\begin{titlepage}

\begin{flushright}
UFIFT-QG-11-09
\end{flushright}

\vspace{1.5cm}

\begin{center}
{\bf Inflationary Scalars Don't Affect Gravitons at One Loop}
\end{center}

\vspace{.5cm}

\begin{center}
Sohyun Park$^{\dagger}$ and R. P. Woodard$^{\ddagger}$
\end{center}

\vspace{.5cm}

\begin{center}
\it{Department of Physics \\
University of Florida \\
Gainesville, FL 32611}
\end{center}

\vspace{1cm}

\begin{center}
ABSTRACT
\end{center}
Primordial inflation results in the production of a vast ensemble of
highly infrared, massless, minimally coupled scalars. We use a
recent fully renormalized computation of the one loop contribution
to the graviton self-energy from these scalars to show that they
have no effect on the propagation of dynamical gravitons. Our
computation motivates a conjecture for the first correction to the
vacuum state wave functional of gravitons. We comment as well on
performing the same analysis for the more interesting contribution
from inflationary gravitons, and on inferring one loop corrections
to the force of gravity.

\vspace{.5cm}

\begin{flushleft}
\end{flushleft}

\vspace{1.5cm}
\begin{flushleft}
$^{\dagger}$ e-mail: spark@phys.ufl.edu \\
$^{\ddagger}$ e-mail: woodard@phys.ufl.edu
\end{flushleft}
\end{titlepage}

\section{Introduction}

Inflation produces a vast ensemble of infrared gravitons and
massless, minimally coupled (MMC) scalars \cite{GPF}. In the theory
of inflationary cosmology these particles are the source of
primordial tensor and scalar perturbations \cite{SMC}, the scalar
component of which has been detected \cite{WMAP}. It is natural to
wonder how this ensemble of quanta changes the propagation of free
particles during inflation.

The effect of inflationary gravitons or scalars on the propagation
of a particular kind of particle is governed by that particle's
one-particle-irreducible (1PI) 2-point function. For scalars this
is the self-mass-squared, $-i M^2(x;x')$; it is the self-energy for
a fermion, $-i [\mbox{}_i \Sigma_j](x;x')$; for a vector it is the
vacuum polarization, $-i[\mbox{}^{\mu} \Pi^{\nu}](x;x')$; and it is
the self-energy for a graviton, $-i[\mbox{}^{\mu\nu}
\Sigma^{\rho\sigma}](x;x')$. One first computes the renormalized
contribution of inflationary gravitons or MMC scalars to the
appropriate 1PI function, then uses this to quantum-correct the
linearized effective field equations. For example, the linearized
effective field equations of a MMC scalar are,
\begin{equation}
\partial_{\mu} \Bigl( \sqrt{-g} g^{\mu\nu} \partial_{\nu} \varphi(x)
\Bigr) - \int \!\! d^4x' \, M^2(x;x') \varphi(x') = 0 \; .
\end{equation}

Many studies of this type have been made over the past decade. The
one loop effects of inflationary scalars have been worked out on
photons, assuming the scalars are charged \cite{SQED}, on fermions,
assuming a Yukawa coupling \cite{Yukawa}, and on other scalars,
assuming either that the scalars have a quartic self-interaction
\cite{scalself}, that they interact electromagnetically \cite{KW1},
or that they interact with fermions \cite{LDW}. The effects of
inflationary gravitons have been worked out for MMC scalars
\cite{KW2} and for massless fermions \cite{MW}.

What happens in each case seems to depend upon whether or not the
highly infrared gravitons and scalars created by inflation can
maintain a significant interaction with the particle in question.
Because neither electromagnetic nor Yukawa charge weakens with
redshift, the effects of inflationary scalars on photons and
fermions is profound: both particles acquire a growing mass
\cite{SQED,Yukawa}. The same is true for MMC scalars with a quartic
self-interaction \cite{scalself}, but the redshift of photons and
fermions means that nothing significant happens to either charged
scalars \cite{KW1} or Yukawa-coupled scalars \cite{LDW}. Because the
spin of infrared gravitons does not redshift, they induce a growing
field strength on fermions \cite{MW}. However, gravitons only interact
with a MMC scalar through the scalar's rapidly redshifting kinetic
energy, and this results in no significant effect\cite{KW2}.

The purpose of this paper is study how inflationary scalars affect
the propagation of free gravitons. We have already computed the
fully renormalized, one loop contribution to the graviton
self-energy from MMC scalars \cite{PW}. That result is summarized in
section 2. In section 3 we solve the linearized effective field
equations at one loop order. Section 4 gives our conclusions.

\section{The Effective Field Equations}

The purpose of this section is to present the effective field
equation which we solve in the next section. We begin by reviewing
some useful facts about the background geometry. We then give our
recently derived result for the one loop MMC scalar contribution
to the graviton self-energy \cite{PW}. The section closes with a
discussion of the Schwinger-Keldysh effective field equations and
how one solves them perturbatively.

\subsection{The Background Geometry}

Our background geometry is the open conformal coordinate submanifold
of $4$-dimensional de Sitter space. A spacetime point $x^{\mu} =
(\eta, x^i)$ takes values in the ranges
\begin{equation}
-\infty < \eta < 0 \qquad {\rm and} \qquad  -\infty < x^i < +\infty \; .
\end{equation}
In these coordinates the invariant element is,
\begin{equation}
ds^2 \equiv g_{\mu\nu} dx^{\mu} dx^{\nu} = a^2 \eta_{\mu\nu}
dx^{\mu} dx^{\nu}\; ,
\end{equation}
where $\eta_{\mu\nu}$ is the Lorentz metric, the scale factor is
$a = -1/H\eta$ and $H$ is the Hubble constant.

It is worth observing that our locally de Sitter geometry should be a
good approximation for primordial inflation. This can be quantified
in terms of the parameter $\epsilon$ which measures how nearly constant
the Hubble parameter is. For a general scale factor, not necessarily de
Sitter, we define $\epsilon$ as,
\begin{equation}
\epsilon \equiv - a^{-1} \frac{d}{d \eta} \Bigl( \frac{d a^{-1}}{d\eta}
\Bigr)^{-1} \; .
\end{equation}
For de Sitter ($a = -1/H\eta$) the result is $\epsilon = 0$. If one
assumes single scalar inflation then the current upper bound on the
tensor-to-scalar ratio \cite{WMAP} implies $\epsilon < 0.014$ at the
time, near the end of inflation, when the largest observable
perturbations experienced horizon crossing \cite{KOW1}. Because
$\epsilon$ is expected to have been even smaller at earlier times,
the de Sitter approximation of $\epsilon = 0$ seems quite
reasonable.

The MMC scalar contribution to the graviton self-energy is de Sitter
invariant and can be expressed using the Sitter length function $y(x;x')$,
\begin{equation}
y(x;x') \equiv a a' H^2 \Bigl[ \Vert \vec{x} \!-\! \vec{x'} \Vert^2
- (\vert \eta \!-\! \eta'\vert \!-\! i \epsilon)^2 \Bigr]\; .
\label{ydef}
\end{equation}
Except for the factor of $i\epsilon$ (whose purpose is to enforce
Feynman boundary conditions) the function $y(x;x')$ is closely
related to the invariant length $\ell(x;x')$ from $x^{\mu}$ to $x'^{\mu}$,
\begin{equation}
y(x;x') = 4 \sin^2\Bigl( \frac12 H \ell(x;x')\Bigr) \; . \label{length}
\end{equation}

With this de Sitter invariant quantity $y(x;x')$, we can form a
convenient basis of de Sitter invariant bi-tensors. Note that
because $y(x;x')$ is de Sitter invariant, so too are covariant
derivatives of it. With the metrics $g_{\mu\nu}(x)$ and
$g_{\mu\nu}(x')$, the first three derivatives of $y(x;x')$ furnish
a convenient basis of de Sitter invariant bi-tensors \cite{KW1},
\begin{eqnarray}
\frac{\partial y(x;x')}{\partial x^{\mu}} & = & H a \Bigl(y
\delta^0_{\mu}
\!+\! 2 a' H \Delta x_{\mu} \Bigr) \; , \label{dydx} \\
\frac{\partial y(x;x')}{\partial x'^{\nu}} & = & H a' \Bigl(y
\delta^0_{\nu}
\!-\! 2 a H \Delta x_{\nu} \Bigr) \; , \label{dydx'} \\
\frac{\partial^2 y(x;x')}{\partial x^{\mu} \partial x'^{\nu}} & = &
H^2 a a' \Bigl(y \delta^0_{\mu} \delta^0_{\nu} \!+\! 2 a' H \Delta
x_{\mu} \delta^0_{\nu} \!-\! 2 a \delta^0_{\mu} H \Delta x_{\nu}
\!-\! 2 \eta_{\mu\nu}\Bigr) \; . \qquad \label{dydxdx'}
\end{eqnarray}
Here and subsequently
$\Delta x_{\mu} \equiv \eta_{\mu\nu} (x \!-\!x')^{\nu}$.

Acting covariant derivatives generates more basis tensors,
for example \cite{KW1},
\begin{eqnarray}
\frac{D^2 y(x;x')}{Dx^{\mu} Dx^{\nu}}
& = & H^2 (2 \!-\!y) g_{\mu\nu}(x) \; , \\
\frac{D^2 y(x;x')}{Dx'^{\mu} Dx'^{\nu}}
& = & H^2 (2 \!-\!y)g_{\mu\nu}(x') \; .
\end{eqnarray}
The contraction of any pair of the basis tensors also produces
more basis tensors \cite{KW1},
\begin{eqnarray}
g^{\mu\nu}(x) \frac{\partial y}{\partial x^{\mu}} \frac{\partial y}{\partial x^{\nu}} & = & H^2 \Bigl(4 y - y^2\Bigr) =
g^{\mu\nu}(x') \frac{\partial y}{
\partial x'^{\mu}} \frac{\partial y}{\partial x'^{\nu}} \; ,
\label{contraction1}\\
g^{\mu\nu}(x) \frac{\partial y}{\partial x^{\nu}} \frac{\partial^2 y}{
\partial x^{\mu} \partial x'^{\sigma}} & = & H^2 (2-y) \frac{\partial y}{
\partial x'^{\sigma}} \; ,
\label{contraction2}\\
g^{\rho\sigma}(x') \frac{\partial y}{\partial x'^{\sigma}}
\frac{\partial^2 y}{\partial x^{\mu} \partial x'^{\rho}} & = & H^2
(2-y)
\frac{\partial y}{\partial x^{\mu}} \; ,
\label{contraction3}\\
g^{\mu\nu}(x) \frac{\partial^2 y}{\partial x^{\mu} \partial
x'^{\rho}} \frac{\partial^2 y}{\partial x^{\nu} \partial
x'^{\sigma}} & = & 4 H^4 g_{\rho\sigma}(x') - H^2 \frac{\partial
y}{\partial x'^{\rho}}
\frac{\partial y}{\partial x'^{\sigma}} \; ,
\label{contraction4}\\
g^{\rho\sigma}(x') \frac{\partial^2 y}{\partial x^{\mu}\partial
x'^{\rho}} \frac{\partial^2 y}{\partial x^{\nu} \partial
x'^{\sigma}} & = & 4 H^4 g_{\mu\nu}(x) - H^2 \frac{\partial
y}{\partial x^{\mu}} \frac{\partial y}{
\partial x^{\nu}} \; .
\label{contraction5}
\end{eqnarray}

Our basis tensors are naturally covariant, but their indices can of
course be raised using the metric at the appropriate point. To save
space in writing this out we define the basis tensors with raised
indices as differentiation with respect to ``covariant''
coordinates,
\begin{eqnarray}
\frac{\partial y}{\partial x_{\mu}} & \equiv &
g^{\mu\nu}(x)
\frac{\partial y}{\partial x^{\nu}} \; , \\
\frac{\partial y}{\partial x'_{\rho}} & \equiv &
g^{\rho\sigma}(x') \frac{\partial y}{\partial x^{\prime
\sigma}} \; , \\
\frac{\partial^2 y}{\partial x_{\mu} \partial x'_{\rho}} & \equiv &
g^{\mu\nu}(x) g^{\rho\sigma}(x')
\frac{\partial^2 y}{\partial x^{\nu} \partial x^{\prime \sigma}} \;
.
\end{eqnarray}

\subsection{The Graviton Self-Energy}

It is simple to infer the unrenormalized one loop scalar 
contribution to the graviton self-energy from the correlator
of two stress tensors at noncoincident points \cite{PNRV}. However, 
an enormous amount of labor is necessary to extract enough 
derivative operators to segregate the ultraviolet divergences onto 
local counterterms, leaving a result which is integrable in the 
$D=4$ effective field equations. This fully renormalized result 
takes the form \cite{PW},
\begin{eqnarray}
\lefteqn{ -i \Bigl[\mbox{}^{\mu\nu} \Sigma^{\rho\sigma}\Bigr](x;x') =
\sqrt{-g(x)} \, \mathcal{P}^{\mu\nu}(x)
\sqrt{-g(x')} \, \mathcal{P}^{\rho\sigma}(x') \Bigl\{
\mathcal{F}_0(y)\Bigr\} } \nonumber \\
& & \hspace{-.5cm} + \sqrt{-g(x)} \,
\mathcal{P}^{\mu\nu}_{\alpha\beta \gamma\delta}(x)
\sqrt{-g(x')} \, \mathcal{P}^{\rho\sigma}_{\kappa\lambda
\theta\phi}(x') \Biggl\{\mathcal{T}^{\alpha\kappa} \mathcal{T}^{\beta\lambda}
\mathcal{T}^{\gamma\theta} \mathcal{T}^{\delta\phi}
\Bigl( \frac{D \!-\! 2}{D \!-\! 3}\Bigr) \mathcal{F}_2(y) \Biggr\} ,
\qquad \label{ansatz}
\end{eqnarray}
where the bi-tensor $\mathcal{T}^{\alpha\kappa}$ is,
\begin{equation}
\mathcal{T}^{\alpha\kappa}(x;x') \equiv -\frac1{2 H^2} \,
\frac{\partial^2 y(x;x')}{\partial x_{\alpha} \partial x'_{\kappa}}
\; . \label{Tak}
\end{equation}
The other quantities in this expression are the spin zero and spin
two projectors, $\mathcal{P}^{\mu\nu}$ and
$\mathcal{P}^{\mu\nu}_{\alpha\beta \gamma\delta}$, respectively, and
their associated structure functions, $\mathcal{F}_0(y)$ and
$\mathcal{F}_2(y)$. We shall devote a paragraph to each.

The two projectors come from expanding the scalar and Weyl curvatures
around de Sitter background,
\begin{eqnarray}
R - D (D \!-\! 1) H^2 & \equiv & \mathcal{P}^{\mu\nu} \kappa h_{\mu\nu}
+ O(\kappa^2 h^2) \; , \label{Rexp} \\
C_{\alpha\beta\gamma\delta}  & \equiv & \mathcal{P}^{\mu\nu}_{\alpha\beta
\gamma\delta} \kappa h_{\mu\nu} + O(\kappa^2 h^2) \; . \label{Cexp}
\end{eqnarray}
From (\ref{Rexp}) we have,
\begin{equation}
\mathcal{P}^{\mu\nu} = D^{\mu} D^{\nu} - g^{\mu\nu}
\Bigl[ D^2 + (D \!-\! 1) H^2\Bigr] \; , \label{P0}
\end{equation}
where $D^{\mu}$ is the covariant derivative operator in de Sitter
background. The more difficult expansion of the Weyl tensor gives,
\begin{eqnarray}
\lefteqn{\mathcal{P}^{\mu\nu}_{\alpha\beta\gamma\delta} =
\mathcal{D}^{\mu\nu}_{\alpha\beta\gamma\delta} + \frac1{D \!-\!2}
\Bigl[ g_{\alpha\delta} \mathcal{D}^{\mu\nu}_{\beta\gamma}
\!-\! g_{\beta\delta} \mathcal{D}^{\mu\nu}_{\alpha\gamma}
\!-\! g_{\alpha\gamma} \mathcal{D}^{\mu\nu}_{\beta\delta}
\!+\! g_{\beta\gamma} \mathcal{D}^{\mu\nu}_{\alpha\delta}
\Bigr] } \nonumber \\
& & \hspace{5cm} + \frac1{(D \!-\! 1) (D \!-\! 2)}
\Bigl[g_{\alpha\gamma} g_{\beta\delta} \!-\!
g_{\alpha\delta} g_{\beta\gamma} \Bigr]
\mathcal{D}^{\mu\nu} \; , \qquad \label{P2}
\end{eqnarray}
where we define,
\begin{eqnarray}
\mathcal{D}^{\mu\nu}_{\alpha\beta\gamma\delta} & \equiv & \frac12 \Bigl[
\delta^{(\mu}_{\alpha} \delta^{\nu)}_{\delta} D_{\gamma} D_{\beta}
\!-\! \delta^{(\mu}_{\beta} \delta^{\nu)}_{\delta} D_{\gamma} D_{\alpha}
\!-\! \delta^{(\mu}_{\alpha} \delta^{\nu)}_{\gamma} D_{\delta} D_{\beta}
\!+\! \delta^{(\mu}_{\beta} \delta^{\nu)}_{\gamma} D_{\delta} D_{\alpha}
\Bigr] \; , \qquad \\
\mathcal{D}^{\mu\nu}_{\beta\delta} & \equiv & g^{\alpha\gamma}
\mathcal{D}^{\mu\nu}_{\alpha\beta\gamma\delta} = \frac12 \Bigl[
\delta^{(\mu}_{\delta} D^{\nu)} D_{\beta} \!-\! \delta^{(\mu}_{\beta}
\delta^{\nu)}_{\delta} D^2 \!-\! g^{\mu\nu} D_{\delta}
D_{\beta} \!+\! \delta^{(\mu}_{\beta} D_{\delta} D^{\nu)}
\Bigr] \; , \qquad \\
\mathcal{D}^{\mu\nu} & \equiv & g^{\alpha\gamma}
g^{\beta\delta} \mathcal{D}^{\mu\nu}_{\alpha\beta\gamma\delta}
= D^{(\mu} D^{\nu)} - g^{\mu\nu} D^2 \; .
\end{eqnarray}

The spin zero structure function is,
\begin{eqnarray}
\lefteqn{\mathcal{F}_{0} = \frac{\kappa^2 H^4}{(4\pi)^4} \Biggl\{
\frac{\square}{H^2} \Biggl[ \frac1{72} \!\times\! \frac{4}{y} \ln\Bigl(
\frac{y}4\Bigr) \Biggr] \!-\! \frac1{12} \!\times\! \frac{4}{y} \ln\Bigl(
\frac{y}{4}\Bigr) \!+\! \frac1{72} \!\times\! \frac{4}{y} \!+\! \frac16
\ln^2\Bigl(\frac{y}{4}\Bigr) }
\nonumber \\
& & \hspace{1.5cm}
+\frac{1}{45} \!\times\! \frac4{4-y}\ln(\frac{y}{4})
-\frac{1}{45} \ln(\frac{y}{4})
+\frac{43}{216} \!\!\times\!\! \frac4{4-y}
-\frac56 \!\times\!\frac{y}{4}\ln(1 - \frac{y}{4})
\nonumber \\
& & \hspace{1.5cm}
+\frac{7}{90} \!\times\!\frac{4}{y}\ln(1 - \frac{y}{4})
-\frac1{20}\ln(1 - \frac{y}{4})
-\frac{7(12 \pi^2 +265)}{540} \!\times\! \frac{y}{4}
\nonumber \\
& & \hspace{1.5cm}
+\frac{84 \pi^2 - 131}{1080}
-\frac13 \!\times\! \frac{y}{4} \ln^2\Bigl(\frac{y}{4}\Bigr)
+ \frac49 \!\times\! \frac{y}{4} \ln\Bigl(\frac{y}{4}\Bigr)
\nonumber \\
& & \hspace{1.5cm}
-\frac1{30} (2 - y) \biggl[ 7\mbox{Li}_2(1-\frac{y}{4})
- 2 \mbox{Li}_2(\frac{y}{4}) +  5\ln(1 - \frac{y}{4})\ln(\frac{y}{4}) \biggr]
\Biggr\} \;. \qquad
\label{F0}
\end{eqnarray}
Here ${\rm Li}_2(z)$ is the dilogarithm function,
\begin{equation}
{\rm Li}_2(z) \equiv -\int_0^z \!\! dt \, \frac{\ln(1 \!-\! t)}{t} =
\sum_{k=1}^{\infty} \frac{z^k}{k^2} \; . \label{dilog}
\end{equation}
The same function also appears in the spin two structure function,
\begin{eqnarray}
\lefteqn{\mathcal{F}_{2} = \frac{\kappa^2 H^4}{(4\pi)^4} \Biggl\{
\frac{\square}{H^2} \Biggl[ \frac1{240} \!\times\! \frac{4}{y}
\ln(\Bigl( \frac{y}4\Bigr) \Biggr] \!+\! \frac3{40} \!\times\!
\frac{4}{y} \ln\Bigl( \frac{y}{4}\Bigr) \!-\! \frac{11}{48} \!\times\!
\frac{4}{y}
+ \frac14 \ln^2\Bigl(\frac{y}{4}\Bigr) }
\nonumber \\
& & \hspace{1cm}
- \frac{119}{60} \ln\Bigl(\frac{y}{4}\Bigr)
+\frac{4096}{(4y-y^2 -8)^4}\Biggl[
\biggl[-\frac{47}{15}\Bigl(\frac{y}{4}\Bigr)^8 + \frac{141}{10}
\Bigl(\frac{y}{4}\Bigr)^7 \nonumber \\
& & \hspace{1cm}
- \frac{2471}{90}\Bigl(\frac{y}{4}\Bigr)^6
+ \frac{34523}{720}\Bigl(\frac{y}{4}\Bigr)^5
-\frac{132749}{1440}\Bigl(\frac{y}{4}\Bigr)^4 +
\frac{38927}{320}\Bigl(\frac{y}{4}\Bigr)^3 \nonumber \\
& & \hspace{1cm}
- \frac{10607}{120} \Bigl(\frac{y}{4}\Bigr)^2
+ \frac{22399}{720}\Bigl(\frac{y}{4}\Bigr)
- \frac{3779}{960} \biggr]\frac{4}{4-y}
+ \biggl[\frac{193}{30}\Bigl(\frac{y}{4}\Bigr)^4
-\frac{131}{10}\Bigl(\frac{y}{4}\Bigr)^3
\nonumber \\
& & \hspace{1cm}
+ \frac{7}{20} \Bigl(\frac{y}{4}\Bigr)^2
+\frac{379}{60} \Bigl(\frac{y}{4}\Bigr)-
\frac{193}{120}\biggr]\ln(2-\frac{y}{2})
+ \biggl[-\frac{14}{15} \Bigl(\frac{y}{4}\Bigr)^5
-\frac{1}{5}\Bigl(\frac{y}{4}\Bigr)^4 \nonumber \\
& & \hspace{1cm} +\frac{19}{2} \Bigl(\frac{y}{4}\Bigr)^3
-\frac{889}{60}\Bigl(\frac{y}{4}\Bigr)^2
+ \frac{143}{20}\Bigl(\frac{y}{4}\Bigr)
-\frac{13}{20} -\frac{7}{60}\Bigl(\frac{4}{y}\Bigr)\biggr]\ln(1
-\frac{y}{4}) \nonumber \\
& &  \hspace{1cm} +\biggl[-\frac{476}{15}\Bigl(\frac{y}{4}\Bigr)^9
+ 160 \Bigl(\frac{y}{4}\Bigr)^8
-\frac{5812}{15} \Bigl(\frac{y}{4}\Bigr)^7
+\frac{8794}{15} \Bigl(\frac{y}{4}\Bigr)^6 \nonumber \\
& & \hspace{1.5cm} -\frac{18271 }{30} \Bigl(\frac{y}{4}\Bigr)^5
+\frac{54499}{120} \Bigl(\frac{y}{4}\Bigr)^4
-\frac{59219}{240} \Bigl(\frac{y}{4}\Bigr)^3
+\frac{1917}{20}\Bigl(\frac{y}{4}\Bigr)^2 \nonumber \\
& & \hspace{1.5cm} -\frac{1951}{80} \Bigl(\frac{y}{4}\Bigr)
+\frac{367}{120}\biggr]\frac{4}{4-y}\ln(\frac{y}{4})
+\biggl[4 \Bigl(\frac{y}{4}\Bigr)^7-12 \Bigl(\frac{y}{4}\Bigr)^6
+20 \Bigl(\frac{y}{4}\Bigr)^5 \nonumber \\
& & \hspace{1.5cm} -20 \Bigl(\frac{y}{4}\Bigr)^4
+15 \Bigl(\frac{y}{4}\Bigr)^3-7 \Bigl(\frac{y}{4}\Bigr)^2
+\Bigl(\frac{y}{4}\Bigr)\biggr]\frac{4-y}{4}\ln^2(\frac{y}{4}) \nonumber \\
& & \hspace{1cm} +\biggl[\frac{367}{30} \Bigl(\frac{y}{4}\Bigr)^4
-\frac{4121}{120}\Bigl(\frac{y}{4}\Bigr)^3
+\frac{237}{16}\Bigl(\frac{y}{4}\Bigr)^2
+\frac{1751}{240}\Bigl(\frac{y}{4}\Bigr)
-\frac{367}{120} \biggr] \ln(\frac{y}{2}) \nonumber \\
& & \hspace{1cm} +\frac{1}{64}(y^2 -8) \Bigl[4(2 - y)
-(4y - y^2)\Bigr]\biggl[\frac{1}{5} \mbox{Li}_2(1-\frac{y}{4})
+\frac{7}{10}\mbox{Li}_2(\frac{y}{4})\biggr] \Biggr]\Biggr\} \; . \nonumber \\
\label{F2}
\end{eqnarray}
Note that these results were derived for Bunch-Davies vacuum, which 
corresponds to a state which is minimum energy in the distant past \cite{PW}.
This is the standard choice for inflationary perturbations, and the
choice we must make in order to compute quantum corrections to the
usual tree order results.

\subsection{The Schwinger-Keldysh Effective Field Equations}

Because the graviton self-energy is the 1PI graviton 2-point
function, it gives the quantum correction to the linearized Einstein
equation,
\begin{equation}
\sqrt{-g} \, \mathcal{D}^{\mu\nu\rho\sigma} h_{\rho\sigma}(x) - \int
\!\! d^4x' \Bigl[ \mbox{}^{\mu\nu} \Sigma^{\rho\sigma}\Bigr](x;x')
h_{\rho\sigma}(x') = \frac12 \kappa \sqrt{-g} \,
T^{\mu\nu}_{\mbox{\tiny lin}}(x) \; , \label{lineqn}
\end{equation}
Here $\mathcal{D}^{\mu\nu\rho\sigma}$ is the Lichnerowicz operator,
specialized to de Sitter background
\begin{eqnarray}
\lefteqn{ \mathcal{D}^{\mu\nu\rho\sigma} \equiv D^{(\rho} g^{\sigma)
(\mu} D^{\nu)} -\frac12 \Bigl[g^{\rho\sigma} D^{\mu} D^{\nu} \!+\!
g^{\mu\nu} D^{\rho} D^{\sigma}\Bigr] } \nonumber \\
& & \hspace{1.5cm} + \frac12 \Bigl[ g^{\mu\nu} g^{\rho\sigma} \!-\!
g^{\mu (\rho} g^{\sigma) \nu} \Bigr] D^2 + (D\!-\!1) \Bigl[\frac12
g^{\mu\nu} g^{\rho\sigma} \!-\! g^{\mu (\rho} g^{\sigma) \nu}\Bigr]
H^2 \; , \qquad \label{Lich}
\end{eqnarray}
and $D^{\mu}$ is the covariant derivative operator in the background
geometry.

Two embarrassments would confront us were we to solve equation
(\ref{lineqn}) using the self-energy of the previous sub-section:
\begin{itemize}
\item{{\it Causality violation} --- the field equation at $x^{\mu}$ involves
the field at points ${x'}^{\mu}$ outside the past light-cone of
$x^{\mu}$; and}
\item{{\it Reality violation} --- the quantum-induced graviton field
would acquire an imaginary part due to the nonzero imaginary part of
the in-out self-energy.}
\end{itemize}
Both features are the result of taking the in-out matrix element of
the operator field equations. This isn't wrong, in fact it is
exactly the right thing to do in the study of asymptotic scattering
problems. However, there is no S-matrix in de Sitter space
\cite{EWAS}, so the more natural problem is to release the universe
in a prepared initial state and then watch it evolve.

The correct effective field equations for releasing the universe in
a prepared initial state are derived by taking the expectation value
of the operator field equations in that state. They are given by the
Schwinger-Keldysh formalism \cite{Schwinger} which, for our problem,
amounts to replacing the in-out self-energy in (\ref{lineqn}) by the
sum of two of the four Schwinger-Keldysh self-energies,
\begin{equation}
\Bigl[\mbox{}^{\mu\nu} \Sigma^{\rho\sigma}\Bigr](x;x')
\longrightarrow \Bigl[\mbox{}^{\mu\nu}
\Sigma^{\rho\sigma}\Bigr]_{\scriptscriptstyle ++}\!\!(x;x') +
\Bigl[\mbox{}^{\mu\nu} \Sigma^{\rho\sigma}\Bigr]_{\scriptscriptstyle
+-}\!\!(x;x') \; .
\end{equation}
At the one loop order we are working $[\mbox{}^{\mu\nu}
\Sigma^{\rho\sigma}]_{\scriptscriptstyle ++}\!(x;x')$ agrees exactly
with the in-out result given in the previous sub-section. To get
$[\mbox{}^{\mu\nu} \Sigma^{\rho\sigma}]_{\scriptscriptstyle
+-}(x;x')$, at this order, one simply adds a minus sign and replaces
the de Sitter length function $y(x;x')$ everywhere with,
\begin{equation}
y(x;x') \longrightarrow y_{\scriptscriptstyle +-}\!(x;x') \equiv H^2
a(\eta) a(\eta') \Bigl[\Vert \vec{x} \!-\! \vec{x}'\Vert^2 - (\eta
\!-\! \eta' \!+\! i \epsilon)^2 \Bigr] \; .
\end{equation}
It will be seen that the $++$ and $+-$ self-energies cancel unless
the point $x^{\prime \mu}$ is on or inside the past light-cone of
$x^{\mu}$. That makes the effective field equation (\ref{lineqn})
causal. When $x^{\prime \mu}$ is on or inside the past light-cone of
$x^{\mu}$ the $+-$ self-energy is the complex conjugate of the $++$
one, which makes the effective field equation (\ref{lineqn}) real.
This also effects a great simplification in the structure functions
because only those terms with branch cuts in $y$ can make nonzero
contributions, for example,
\begin{equation}
\ln( y_{\scriptscriptstyle ++}) - \ln(y_{\scriptscriptstyle +-}) =
2\pi i \theta\Bigl( \eta \!-\! \eta' - \Vert \vec{x} \!-\! \vec{x}'
\Vert \Bigr) \; .
\end{equation}

\subsection{Perturbative Solution}

Because we only know the self-energy at one loop order, all we can
do is to solve (\ref{lineqn}) perturbatively by expanding the
graviton field and the self-energy in powers of $\kappa^2$,
\begin{equation}
h_{\mu\nu}(x) = h^{(0)}_{\mu\nu}(x) + \kappa^2 h^{(1)}_{\mu\nu}(x) +
O(\kappa^4) \; .
\end{equation}
Of course $h^{(0)}_{\mu\nu}(x)$ obeys the classical, linearized
Einstein equation. Given this solution, the corresponding one loop
correction is defined by the equation,
\begin{equation}
\sqrt{-g(x)} \, \mathcal{D}^{\mu\nu\rho\sigma} \kappa^2
h^{(1)}_{\rho\sigma}(x) = \int \!\! d^4x' \, \Bigl[\mbox{}^{\mu\nu}
\Sigma^{\rho\sigma}\Bigr](x;x') h^{(0)}_{\rho\sigma}(x') \; .
\label{oneloop}
\end{equation}

The classical solution for a dynamical graviton of wave vector
$\vec{k}$ is \cite{TW3},
\begin{equation}
h^{(0)}_{\rho\sigma}(x) = \epsilon_{\rho\sigma}(\vec{k}) u(\eta,k)
e^{i \vec{k} \cdot \vec{x}} \; , \label{dyn1}
\end{equation}
where the tree order mode function is,
\begin{equation}
u(\eta,k) = \frac{H}{\sqrt{2 k^3}} \Bigl[1 - \frac{i k}{H a}\Bigr]
\exp\Bigl[ \frac{ik}{H a}\Bigr] \; , \label{dyn2}
\end{equation}
and the polarization tensor obeys all the same relations as in flat
space,
\begin{equation}
0 = \epsilon_{0\mu} = k_i \epsilon_{ij} = \epsilon_{jj} \quad {\rm
and} \quad \epsilon_{ij} \epsilon_{ij}^* = 1 \; . \label{dyn3}
\end{equation}

\section{Computing the One Loop Source}

The point of this section is to evaluate the one loop source term on
the right hand side of equation (\ref{oneloop}) for a dynamical
graviton (\ref{dyn1}-\ref{dyn3}). We begin by drawing inspiration
from what happens in the flat space limit. Our de Sitter analysis
commences by partially integrating the projectors. This results in
considerable simplification but the plethora of indices is still
problematic. To effect further simplification we extract and
partially integrate another d'Alembertian, whereupon the $x^{\mu}$
projector can be acted on the residual structure function to
eliminate four contractions. At this point we digress to derive some
important identities concerning covariant derivatives of the Weyl
tensor. The final reduction reveals zero net result.

\subsection{The Flat Space Limit}

The one loop contribution to the graviton self-energy from MMC
scalars in a flat background was first computed by `t Hooft and
Veltman in 1974 \cite{HV}. When renormalized and expressed in
position space using the Schwinger-Keldysh formalism the result
takes the form \cite{FW},
\begin{equation}
\Bigl[\mbox{}^{\mu\nu} \Sigma^{\rho\sigma}_{\rm flat}\Bigr](x;x') =
\Pi^{\mu\nu} \Pi^{\rho\sigma} F_0(\Delta x^2) + \Bigl[ \Pi^{\mu
(\rho} \Pi^{\sigma) \nu} \!-\! \frac13 \Pi^{\mu\nu}
\Pi^{\rho\sigma}\Bigr] F_2(\Delta x^2) \; . \label{flatSigma}
\end{equation}
Here $\Pi^{\mu\nu} \equiv \partial^{\mu} \partial^{\nu} -
\eta^{\mu\nu} \partial^2$ and the two structure functions are,
\begin{eqnarray}
F_0(\Delta x^2) & = & \frac{i \kappa^2}{(4 \pi)^4}
\frac{\partial^2}{9} \Biggl[ \frac{\ln(\mu^2 \Delta
x^2_{\scriptscriptstyle ++})}{\Delta x^2_{\scriptscriptstyle ++}} -
\frac{\ln(\mu^2 \Delta x^2_{\scriptscriptstyle +-})}{\Delta
x^2_{\scriptscriptstyle +-}} \Biggr] \; , \qquad \label{flatF0} \\
F_2(\Delta x^2) & = & \frac{i \kappa^2}{(4 \pi)^4}
\frac{\partial^2}{60} \Biggl[ \frac{\ln(\mu^2 \Delta
x^2_{\scriptscriptstyle ++})}{\Delta x^2_{\scriptscriptstyle ++}} -
\frac{\ln(\mu^2 \Delta x^2_{\scriptscriptstyle +-})}{\Delta
x^2_{\scriptscriptstyle +-}} \Biggr] \; \qquad \label{flatF2}
\end{eqnarray}
The two coordinate intervals are,
\begin{eqnarray}
\Delta x^2_{\scriptscriptstyle ++} & \equiv & \Bigl\Vert \vec{x}
\!-\! \vec{x}' \Bigr\Vert^2 - \Bigl( \vert x^0 \!-\! {x'}^0 \vert
\!-\! i \epsilon\Bigr)^2 \; , \qquad \\
\Delta x^2_{\scriptscriptstyle +-} & \equiv & \Bigl\Vert \vec{x}
\!-\! \vec{x}' \Bigr\Vert^2 - \Bigl( x^0 \!-\! {x'}^0 \!+\! i
\epsilon\Bigr)^2 \; . \qquad
\end{eqnarray}
Of course this same form follows from taking the flat space limit of
the de Sitter result summarized in the previous section.

In flat space, the mode function for a plane wave graviton with wave
vector $\vec{k}$ is,
\begin{equation}
h^{\rm flat}_{\mu\nu}(x) = \epsilon_{\rho\sigma}(\vec{k})
\frac1{\sqrt{2 k}} \, e^{-i k x^0 + i \vec{k} \cdot \vec{x}} \; .
\end{equation}
The one loop correction to this (from MMC scalars) is sourced by,
\begin{equation}
\Bigl( {\rm Source}\Bigr)^{\mu\nu}(x) = \int \!\! dx^4x' \,
\Bigl[\mbox{}^{\mu\nu} \Sigma_{\rm flat}^{\rho\sigma}\Bigr](x;x')
h^{\rm flat}_{\rho\sigma}(x') \; . \label{source}
\end{equation}
It might seem natural to extract the various derivatives with
respect to $x^{\mu}$ from the integration, for example,
\begin{eqnarray}
\lefteqn{\int \!\! d^4x' \, \Pi^{\mu\nu} \Pi^{\rho\sigma} F_0(\Delta
x^2) \times h^{\rm flat}_{\rho\sigma}(x') } \nonumber \\
& & \hspace{-.2cm} = \frac{i \kappa^2}{(4\pi)^4} \Pi^{\mu\nu}
\Pi^{\rho\sigma} \frac{\partial^2}{9} \int \!\! d^4x' \, \Biggl[
\frac{\ln(\mu^2 \Delta x^2_{\scriptscriptstyle ++})}{\Delta
x^2_{\scriptscriptstyle ++}} - \frac{\ln(\mu^2 \Delta
x^2_{\scriptscriptstyle +-})}{\Delta x^2_{\scriptscriptstyle +-}}
\Biggr] \times h^{\rm flat}_{\rho\sigma}(x') \; . \qquad
\label{complex}
\end{eqnarray}
That would reduce the source (\ref{source}) to a tedious set of
integrations, followed by some equally tedious differentiations.

The point of this sub-section is that a more efficient strategy is
to first convert all the $x^{\mu}$ derivatives to ${x'}^{\mu}$
derivatives --- which can be done because they act on functions of
$\Delta x^2$. Then ignore surface terms and partially integrate the
${x'}^{\mu}$ derivatives to act upon $h^{\rm
flat}_{\rho\sigma}(x')$. For example, doing this for the spin zero
contribution (\ref{complex}) gives,
\begin{eqnarray}
\lefteqn{\int \!\! d^4x' \, \Pi^{\mu\nu} \Pi^{\rho\sigma} F_0(\Delta
x^2) \times h^{\rm flat}_{\rho\sigma}(x') } \nonumber \\
& & \hspace{-.3cm} \longrightarrow \frac{i \kappa^2}{(4\pi)^4} \int
\!\! d^4x' \, \Biggl[ \frac{\ln(\mu^2 \Delta x^2_{\scriptscriptstyle
++})}{\Delta x^2_{\scriptscriptstyle ++}} - \frac{\ln(\mu^2 \Delta
x^2_{\scriptscriptstyle +-})}{\Delta x^2_{\scriptscriptstyle +-}}
\Biggr] \times \frac{{\partial'}^2}{9} {\Pi'}^{\mu\nu}
{\Pi'}^{\rho\sigma} h^{\rm flat}_{\rho\sigma}(x') \; . \qquad
\label{simple}
\end{eqnarray}
Because the graviton mode function is both transverse and traceless,
we have ${\Pi'}^{\rho\sigma} h^{\rm flat}_{\rho\sigma}(x') = 0$. The
spin two contribution is only a little more complicated,
\begin{eqnarray}
\lefteqn{\int \!\! d^4x' \, \Bigl[\Pi^{\mu (\rho} \Pi^{\sigma) \nu}
- \frac13 \Pi^{\mu\nu} \Pi^{\rho\sigma} \Bigr] F_2(\Delta x^2)
\times h^{\rm flat}_{\rho\sigma}(x') } \nonumber \\
& & \hspace{1cm} \longrightarrow \frac{i \kappa^2}{(4\pi)^4} \int
\!\! d^4x' \, \Biggl[ \frac{\ln(\mu^2 \Delta x^2_{\scriptscriptstyle
++})}{\Delta x^2_{\scriptscriptstyle ++}} - \frac{\ln(\mu^2 \Delta
x^2_{\scriptscriptstyle +-})}{\Delta x^2_{\scriptscriptstyle +-}}
\Biggr] \times \frac{{\partial'}^6}{60} h_{\rm flat}^{\mu\nu}(x') \;
. \qquad \label{alsosimple}
\end{eqnarray}
This also vanishes because ${\partial'}^2 h^{\rm
flat}_{\rho\sigma}(x') = 0$.

In expressions (\ref{simple}) and (\ref{alsosimple}) we have
employed a rightarrow, rather than an equals sign, because the
surface terms produce by partial integration were ignored. There are
no surface terms at spatial infinity in the Schwinger-Keldysh
formalism because the $++$ and $+-$ terms cancel for spacelike
separation. The $++$ and $+-$ contributions also cancel when ${x'}^0
> x^0$, so there are no future surface terms. However, there are
nonzero contributions from the initial value surface.\footnote{For a
two loop example, see \cite{TW5}.} We assume that all such
contributions are absorbed into perturbative corrections to the
initial state, such as has recently been worked out for a MMC scalar
with quartic self-interaction \cite{KOW2}.

\subsection{Partial Integration}

We now start to evaluate the one loop source term (\ref{oneloop}) 
for a dynamical graviton,
\begin{eqnarray}
\lefteqn{ \int \!\! d^4x' \, \Bigl[\mbox{}^{\mu\nu}
\Sigma^{\rho\sigma}\Bigr](x;x') h^{(0)}_{\rho\sigma}(x') }
\nonumber \\
& & \hspace{-.7cm} =
i\!\!\int \!\!  d^4x' \!
\sqrt{\!-\!g(x)} \, \mathcal{P}^{\mu\nu}(x)
\sqrt{\!-\!g(x')} \, \mathcal{P}^{\rho\sigma}(x') \Bigl\{
\mathcal{F}_0\Bigr\} 
h^{(0)}_{\rho\sigma}(x') 
\nonumber \\
& & \hspace{-.7cm} +
2i\!\!\int \!\! d^4x' \!
\sqrt{\!-\!g(x)} \,
\mathcal{P}^{\mu\nu}_{\alpha\beta \gamma\delta}(x)
\sqrt{\!-\!g(x')} \, \mathcal{P}^{\rho\sigma}_{\kappa\lambda
\theta\phi}(x') \Biggl\{\! \mathcal{T}^{\alpha\kappa} 
\mathcal{T}^{\beta\lambda} \mathcal{T}^{\gamma\theta} \mathcal{T}^{\delta\phi}
\mathcal{F}_2 \! \Biggr\}
h^{(0)}_{\rho\sigma}(x') \; . \qquad
\label{oneloop_step0}
\end{eqnarray}
In this expression and henceforth we simply write ``$\mathcal{F}_0$'' 
and ``$\mathcal{F}_2$'' to stand for the full Schwinger-Keldysh 
expressions,
\begin{equation}
\mathcal{F}_0 \equiv \mathcal{F}_0(y_{\scriptscriptstyle ++}) 
- \mathcal{F}_0(y_{\scriptscriptstyle +-}) \qquad , \qquad
\mathcal{F}_2 \equiv \mathcal{F}_2(y_{\scriptscriptstyle ++}) 
- \mathcal{F}_2(y_{\scriptscriptstyle +-}) \; .
\end{equation}
The integral (\ref{oneloop_step0}) can be simplified in two steps. 
First, the projectors $\mathcal{P}^{\mu\nu}(x)$ and
$\mathcal{P}^{\mu\nu}_{\alpha\beta\gamma\delta}(x)$, which act on a 
function of $x^{\mu}$, can be pulled outside the integration over 
${x'}^{\mu}$. Second, the projectors $\mathcal{P}^{\rho\sigma}(x')$ 
and $\mathcal{P}^{\rho\sigma}_{\kappa\lambda\theta\phi}(x')$, which 
act on $x^{\prime \mu}$, can be partially integrated to act on the
graviton wave function $h^{(0)}_{\rho\sigma}(x')$. After these two 
steps, the integral (\ref{oneloop_step0}) becomes,
\begin{eqnarray}
\lefteqn{ \int \!\! d^4x' \, \Bigl[\mbox{}^{\mu\nu}
\Sigma^{\rho\sigma}\Bigr](x;x') h^{(0)}_{\rho\sigma}(x') }
\nonumber \\
& & \hspace{-.5cm} =
i\sqrt{\!-\!g(x)} \, \mathcal{P}^{\mu\nu}(x) \!
\int \!\!  d^4x' \!
\sqrt{\!-\!g(x')} \,  
\mathcal{F}_0 
\Bigl\{\mathcal{P}^{\rho\sigma}(x')
h^{(0)}_{\rho\sigma}(x') \Bigr\}
\nonumber \\
& & \hspace{-.5cm} +
2i \sqrt{\!-\!g(x)} \,
\mathcal{P}^{\mu\nu}_{\alpha\beta \gamma\delta}(x) \!\!
\int \!\! d^4x' \!
\sqrt{\!-\!g(x')} \, \mathcal{T}^{\alpha\kappa} \mathcal{T}^{\beta\lambda}
\mathcal{T}^{\gamma\theta} \mathcal{T}^{\delta\phi}
\mathcal{F}_2
\Biggl\{\! \mathcal{P}^{\rho\sigma}_{\kappa\lambda\theta\phi}(x')
h^{(0)}_{\rho\sigma}(x') \! \Biggr\} . \qquad
\label{oneloop_step1}
\end{eqnarray}

Note that the spin zero term drops out due to the tranversality and 
tracelessness of the dynamical graviton, $h^{(0)}_{\rho\sigma}$:
\begin{equation}
\mathcal{P}^{\rho\sigma} h^{(0)}_{\rho\sigma} = \Bigl\{
D^{\rho} D^{\sigma} - 
\Bigl[ D^2 + (D \!-\! 1) H^2 \Bigr]g^{\rho\sigma}\Bigr\} h^{(0)}_{\rho\sigma} = 0 \; .
\end{equation}
Thus we only have the spin two term, which gives the linearized Weyl tensor,
\begin{equation}
\mathcal{P}^{\rho\sigma}_{\kappa\lambda\theta\phi}(x')
h^{(0)}_{\rho\sigma}(x') = \delta C_{\kappa\lambda\theta\phi}(x') \; .
\end{equation}
The one loop source term then reduces to the integral, 
\begin{eqnarray}
\lefteqn{ \int \!\! d^4x' \, \Bigl[\mbox{}^{\mu\nu}
\Sigma^{\rho\sigma}\Bigr](x;x') h^{(0)}_{\rho\sigma}(x') }
\nonumber \\
& & \hspace{.5cm} =
2i\sqrt{\!-\!g(x)} \,
\mathcal{P}^{\mu\nu}_{\alpha\beta \gamma\delta}(x) \!
\int \!\! d^4x' \!
\sqrt{\!-\!g(x')} \, \mathcal{T}^{\alpha\kappa} \mathcal{T}^{\beta\lambda}
\mathcal{T}^{\gamma\theta} \mathcal{T}^{\delta\phi}
\mathcal{F}_2 
\delta C_{\kappa\lambda\theta\phi}(x') \; . \qquad \label{oneloop_step2}
\end{eqnarray}

\subsection{Extracting Another d'Alembertian}

A challenge to evaluating expression (\ref{oneloop_step2})
is the complicated tensor structure of the external projector
$\mathcal{P}^{\mu\nu}_{\alpha\beta\gamma\delta}(x)$ acting on the
internal factors of $\mathcal{T}^{\alpha\kappa} \cdots \mathcal{F}_2$. 
Recall from the flat space limit that all of this was converted to
derivatives with respect to $x^{\prime \mu}$ and then partially
integrated onto the graviton wave function to give zero. To follow
this on de Sitter we must make the structure function more convergent
by extracting a factor of $\square'$ and then partially integrating
it onto the graviton wave function. After this the external projector
can be acted, which eliminates four indices, and a final further 
partial integration can be performed.

The first step is extracting the extra d'Alembertian,
\begin{equation}
\mathcal{F}_2 = \frac{\square'}{H^2} \widehat{\mathcal{F}}_2 \;.
\end{equation}
We next commute the $\square'$ through the factor of
$\mathcal{T}^{\alpha\kappa} \mathcal{T}^{\beta\lambda}
\mathcal{T}^{\gamma\theta} \mathcal{T}^{\delta\phi}$:
\begin{eqnarray}
\lefteqn{ \mathcal{T}^{\alpha\kappa}\mathcal{T}^{\beta\lambda}
\mathcal{T}^{\gamma\theta} \mathcal{T}^{\delta\phi} 
\frac{\square'}{H^2} \widehat{\mathcal{F}}_{2} =
\Bigl(\frac{\square'}{H^2} \!+\! 4\Bigr) \bigg[\mathcal{T}^{\alpha\kappa}
\mathcal{T}^{\beta\lambda} \mathcal{T}^{\gamma\theta} 
\mathcal{T}^{\delta\phi} \widehat{\mathcal{F}}_2 \bigg] } \nonumber \\
& & - \frac{1}{H^2}\widehat{\mathcal{F}'_{2}}\Biggl\{
\frac{\partial y}{\partial x_{\alpha}}\frac{\partial y}{\partial x'_{\kappa}}
\mathcal{T}^{\beta\lambda}
\mathcal{T}^{\gamma\theta} \mathcal{T}^{\delta\phi}
+ \cdots + \mathcal{T}^{\alpha\kappa}\mathcal{T}^{\beta\lambda}
\mathcal{T}^{\gamma\theta}
\frac{\partial y}{\partial x_{\delta}}\frac{\partial y}{\partial x'_{\phi}}
\Biggl\} 
\nonumber \\
& & -\frac{1}{2 H^2}\widehat{\mathcal{F}_{2}}\Biggl\{ g^{\alpha\beta}
\frac{\partial y}{\partial x'_{\kappa}}\frac{\partial y}{\partial x'_{\lambda}}
\mathcal{T}^{\gamma\theta} \mathcal{T}^{\delta\phi}
+ g^{\alpha\gamma}
\frac{\partial y}{\partial x'_{\kappa}}\frac{\partial y}{\partial x'_{\theta}}
\mathcal{T}^{\beta\lambda}\mathcal{T}^{\delta\phi}
\nonumber \\
& & \hspace{2.5cm} + g^{\alpha\delta}
\frac{\partial y}{\partial x'_{\kappa}}\frac{\partial y}{\partial x'_{\phi}}
\mathcal{T}^{\beta\lambda}\mathcal{T}^{\gamma\theta}
+ g^{\beta\gamma}
\frac{\partial y}{\partial x'_{\lambda}}\frac{\partial y}{\partial x'_{\theta}}
\mathcal{T}^{\alpha\kappa}\mathcal{T}^{\delta\phi}
\nonumber \\
& & \hspace{3.5cm} + g^{\beta\delta}
\frac{\partial y}{\partial x'_{\lambda}}\frac{\partial y}{\partial x'_{\phi}}
\mathcal{T}^{\alpha\kappa}\mathcal{T}^{\gamma\theta}
+ g^{\gamma\delta}
\frac{\partial y}{\partial x'_{\theta}}\frac{\partial y}{\partial x'_{\phi}}
\mathcal{T}^{\alpha\kappa}\mathcal{T}^{\beta\lambda}
\Biggr\} \; . \qquad \label{commute}
\end{eqnarray}
Exploiting the tracelessness of the Weyl tensor on any two indices,
and its antisymmetry on the first two and last two indices, gives,
\begin{eqnarray}
\lefteqn{
{P}^{\mu\nu}_{\alpha\beta \gamma\delta}
\mathcal{T}^{\alpha\kappa}\mathcal{T}^{\beta\lambda}
\mathcal{T}^{\gamma\theta}
\mathcal{T}^{\delta\phi}\frac{\square'}{H^2} \widehat{\mathcal{F}}_{2}
\delta C_{\kappa\lambda\theta\phi}
= {P}^{\mu\nu}_{\alpha\beta \gamma\delta}
\frac{\square'}{H^2}
\bigg[\widehat{\mathcal{F}}_{2} \mathcal{T}^{\alpha\kappa}
\mathcal{T}^{\beta\lambda}
\mathcal{T}^{\gamma\theta}\mathcal{T}^{\delta\phi}\bigg]
\delta C_{\kappa\lambda\theta\phi}
}
\nonumber \\
& &
\hspace{-.5cm} = {P}^{\mu\nu}_{\alpha\beta \gamma\delta}
\Biggl\{
4 \widehat{\mathcal{F}}_{2} \mathcal{T}^{\alpha\kappa}
\mathcal{T}^{\beta\lambda}
\mathcal{T}^{\gamma\theta}\mathcal{T}^{\delta\phi} 
- \frac{4}{H^2} \widehat{\mathcal{F}}'_{2}
\frac{\partial y}{\partial x_{\alpha}}\frac{\partial y}{\partial x'_{\kappa}}
\mathcal{T}^{\beta\lambda}
\mathcal{T}^{\gamma\theta}\mathcal{T}^{\delta\phi}
\Biggr\}
\delta C_{\kappa\lambda\theta\phi} \;.
\qquad
\label{sim_commute}
\end{eqnarray}
For the first term of (\ref{sim_commute}) we can partially integrate the $\square'$ onto the linearized Weyl tensor. Then the one loop source term becomes
\begin{eqnarray}
\lefteqn{ \int \!\! d^4x' \, \Bigl[\mbox{}^{\mu\nu}
\Sigma^{\rho\sigma}\Bigr](x;x') h^{(0)}_{\rho\sigma}(x') }
\nonumber \\
& & \hspace{-.5cm} =
2i\sqrt{\!-\!g(x)} \,
\mathcal{P}^{\mu\nu}_{\alpha\beta \gamma\delta}(x) \!
\int \!\! d^4x' \!\sqrt{\!-\!g(x')} \,
\Bigg\{
\mathcal{T}^{\alpha\kappa} \mathcal{T}^{\beta\lambda}
\mathcal{T}^{\gamma\theta} \mathcal{T}^{\delta\phi}
\widehat{\mathcal{F}}_2
\frac{\square'}{H^2}
\delta C_{\kappa\lambda\theta\phi}(x') 
\nonumber \\
& &
+ \bigg[4 \widehat{\mathcal{F}}_{2}
\mathcal{T}^{\alpha\kappa}\mathcal{T}^{\beta\lambda}
\mathcal{T}^{\gamma\theta}\mathcal{T}^{\delta\phi}
- \frac{4}{H^2} \widehat{\mathcal{F}}'_{2}
\frac{\partial y}{\partial x_{\alpha}}\frac{\partial y}{\partial x'_{\kappa}}
\mathcal{T}^{\beta\lambda}
\mathcal{T}^{\gamma\theta}\mathcal{T}^{\delta\phi} \bigg]
\delta C_{\kappa\lambda\theta\phi}(x')
\Biggr\}
\;.
\label{oneloop_commute}
\end{eqnarray}
This sets the stage for acting the outer projector.

\subsection{Derivatives of the Weyl Tensor}

At this point it is useful to make a short digression on the
covariant derivatives of the Weyl tensor. In this sub-section
we use $g_{\mu\nu}$ for the full metric, not the de Sitter
background. All curvatures are similarly for the full metric.

The Bianchi identity tells us,
\begin{equation}
D_{\epsilon} R_{\alpha\beta\gamma\delta} +
D_{\gamma} R_{\alpha\beta\delta\epsilon} +
D_{\delta} R_{\alpha\beta\epsilon\gamma} = 0 \; . \label{Bianchi}
\end{equation}
If the stress-energy vanishes, all solutions to the Einstein equation
obey,
\begin{equation}
R_{\mu\nu} - \frac12 g_{\mu\nu} R = - 3H^2 g_{\mu\nu} \qquad
\Longrightarrow \qquad R_{\mu\nu} = 3 H^2 g_{\mu\nu} \; . \label{Einstein}
\end{equation}
In $D = 3 + 1$ the Weyl tensor can be expressed in terms of the other
curvatures as,
\begin{equation}
C_{\alpha\beta\gamma\delta} = R_{\alpha\beta\gamma\delta} - \frac12
\Bigl( g_{\alpha\gamma} R_{\beta\delta} - g_{\gamma\beta} R_{\delta\alpha} +
g_{\beta\delta} R_{\alpha\gamma} - g_{\delta\alpha} R_{\gamma\beta} \Bigr) +
\frac16 \Bigl( g_{\alpha\gamma} g_{\beta\delta} -
g_{\alpha\delta} g_{\beta\gamma} \Bigr) R \; . \label{Weyl}
\end{equation}

Now note that the covariant derivative of the metric vanishes. Substituting
(\ref{Einstein}) in (\ref{Weyl}) implies,
\begin{equation}
D_{\epsilon} C_{\alpha\beta\gamma\delta} =
D_{\epsilon} R_{\alpha\beta\gamma\delta} \; . \label{RtoC}
\end{equation}
Combining this relation into (\ref{Bianchi}) gives,
\begin{equation}
D_{\epsilon} C_{\alpha\beta\gamma\delta} +
D_{\gamma} C_{\alpha\beta\delta\epsilon} +
D_{\delta} C_{\alpha\beta\epsilon\gamma} = 0 \; . \label{DCID}
\end{equation}
Our first key identity derives from contracting $\alpha$ into
$\epsilon$, and exploiting the tracelessness of the Weyl tensor,
\begin{equation}
D^{\alpha} C_{\alpha\beta\gamma\delta} = 0 \; . \label{fullID1}
\end{equation}
Our second identity derives from contracting $D^{\epsilon}$ into
relation (\ref{DCID}), commuting derivatives and then using
relation (\ref{fullID1}),
\begin{eqnarray}
\square C_{\alpha\beta\gamma\delta} &=& -D_{\rho} D_{\gamma}
C_{\alpha\beta\delta}^{~~~~ \rho} + D_{\rho} D_{\delta}
C_{\alpha\beta\gamma}^{~~~~ \rho} \; ,  
\\ 
&=& 6 H^2 C_{\alpha\beta\gamma\delta} -
R^{\rho ~~~ \sigma}_{~ \alpha\gamma} C_{\rho\beta\delta\sigma} +
R^{\rho ~~~ \sigma}_{~ \gamma\beta} C_{\rho\delta\alpha\sigma} 
\nonumber \\
& & - R^{\rho ~~~ \sigma}_{~ \beta\delta} C_{\rho\alpha\gamma\sigma} 
+ R^{\rho ~~~ \sigma}_{~ \delta\alpha} C_{\rho\gamma\beta\sigma} 
- R^{\rho\sigma}_{~~ \gamma\delta} C_{\alpha\beta\rho\sigma}\; .
\label{fullID2}
\end{eqnarray}

Relations (\ref{fullID1}) and (\ref{fullID2}) hold, to all orders in
the graviton field, for any solution to the source-free Einstein
equations. Taking the first order in the graviton field amounts to
just replacing the full Weyl tensor by the linearized Weyl $\delta
C_{\alpha\beta\gamma\delta}$ we have been using, replacing the full
covariant derivative operators by the covariant derivatives in de
Sitter background and replacing the full Riemann tensor by its de
Sitter limit. When these things are done the two identities become,
\begin{eqnarray}
D^{\alpha} \delta C_{\alpha\beta\gamma\delta} & = & 0 + O(h^2)
\; , \qquad \label{keyID1} \\
\square \delta C_{\alpha\beta\gamma\delta} & = & 6 H^2 \delta
C_{\alpha\beta\gamma\delta} + O(h^2) \; . \qquad \label{keyID2}
\end{eqnarray}
Note also that if the stress-energy had been nonzero the right hand
sides of relations (\ref{keyID1}) and (\ref{keyID2}) would have
contained simple combinations of derivatives of the stress tensor.

\subsection{The Final Reduction}

We are now ready to act the outer projector on the remaining terms,
\begin{eqnarray}
\lefteqn{ \int \!\! d^4x' \, \Bigl[\mbox{}^{\mu\nu}
\Sigma^{\rho\sigma}\Bigr](x;x') h^{(0)}_{\rho\sigma}(x')
= 2i\sqrt{\!-\!g(x)} \,
\!
\int \!\! d^4x' \!\sqrt{\!-\!g(x')} \,
\delta C_{\kappa\lambda\theta\phi}(x')
}
\nonumber \\
& & \hspace{-.5cm} 
\Bigg\{
\mathcal{P}^{\mu\nu}_{\alpha\beta \gamma\delta}(x) 
\bigg[10 \widehat{\mathcal{F}}_{2}
\mathcal{T}^{\alpha\kappa}\mathcal{T}^{\beta\lambda}
\mathcal{T}^{\gamma\theta}\mathcal{T}^{\delta\phi}
- \frac{4}{H^2} \widehat{\mathcal{F}}'_{2}
\frac{\partial y}{\partial x_{\alpha}}\frac{\partial y}{\partial x'_{\kappa}}
\mathcal{T}^{\beta\lambda}
\mathcal{T}^{\gamma\theta}\mathcal{T}^{\delta\phi} \bigg]
\Biggr\}
\;.
\label{oneloop_act_outerP}
\end{eqnarray}
The second line of this expression is quite complicated by itself,
but it is greatly simplified when contracted into the linearized Weyl
tensor,
\begin{eqnarray}
\lefteqn{\delta C_{\kappa\lambda\theta\phi}(x')
\mathcal{P}^{\mu\nu}_{\alpha\beta \gamma\delta}(x) 
\bigg[10 \widehat{\mathcal{F}}_{2}
\mathcal{T}^{\alpha\kappa}\mathcal{T}^{\beta\lambda}
\mathcal{T}^{\gamma\theta}\mathcal{T}^{\delta\phi}
- \frac{4}{H^2} \widehat{\mathcal{F}}'_{2}
\frac{\partial y}{\partial x_{\alpha}}\frac{\partial y}{\partial x'_{\kappa}}
\mathcal{T}^{\beta\lambda}
\mathcal{T}^{\gamma\theta}\mathcal{T}^{\delta\phi} \bigg]
}
\nonumber \\
& & 
=
\delta C_{\kappa\lambda\theta\phi}(x')\Biggl\{\frac{\partial y}{\partial x'_{\kappa}}\frac{\partial y}{\partial x'_{\theta}}
\mathcal{T}^{\lambda(\mu}\mathcal{T}^{\nu)\phi} f_1(y)
+
\frac{\partial y}{\partial x'_{\kappa}}\frac{\partial y}{\partial x'_{\phi}}
\mathcal{T}^{\lambda(\mu}\mathcal{T}^{\nu)\theta} f_2(y)
\nonumber \\
& & \hspace{2.5cm} 
+
\frac{\partial y}{\partial x'_{\lambda}}\frac{\partial y}{\partial x'_{\theta}}
\mathcal{T}^{\kappa(\mu}\mathcal{T}^{\nu)\phi} f_3(y)
+
\frac{\partial y}{\partial x'_{\lambda}}\frac{\partial y}{\partial x'_{\phi}}
\mathcal{T}^{\kappa(\mu}\mathcal{T}^{\nu)\theta} f_4(y)
\Biggr\}
\;.
\quad
\label{act_outerP}
\end{eqnarray}
Here the functions $f_i(y)$ are,
\begin{eqnarray}
f_1 &\!\!\!\!=\!\!\!\!& 
-125 \widehat{\mathcal{F}}_{2} 
\!+\! 115(2\!-\!y) \widehat{\mathcal{F}}'_{2} 
\!-\!(68-116y +29y^2) \widehat{\mathcal{F}}''_{2} 
\!-\!2(2\!-\!y)(4y\!-\!y^2) \widehat{\mathcal{F}}'''_{2} 
\nonumber \\
f_2 &\!\!\!\!=\!\!\!\!& 
-\frac{75}{2} \widehat{\mathcal{F}}_{2} 
\!+\! \frac{69}{2}(2\!-\!y) \widehat{\mathcal{F}}'_{2} 
\!-\!(28-44y +11y^2) \widehat{\mathcal{F}}''_{2} 
\!-\!(2\!-\!y)(4y\!-\!y^2) \widehat{\mathcal{F}}'''_{2} 
\nonumber \\
f_3 &\!\!\!\!=\!\!\!\!& 
-\frac{85}{2} \widehat{\mathcal{F}}_{2} 
\!+\! \frac{15}{2}(2\!-\!y) \widehat{\mathcal{F}}'_{2} 
\nonumber \\
f_4 &\!\!\!\!=\!\!\!\!& 
-5 \widehat{\mathcal{F}}_{2} 
\!-\! 13(2\!-\!y) \widehat{\mathcal{F}}'_{2} 
\!-\!\frac{5}{2}(4y\!-\!y^2) \widehat{\mathcal{F}}''_{2}
\end{eqnarray}
Changing the dummy indices in (\ref{act_outerP}) gives,
\begin{eqnarray}
\lefteqn{\delta C_{\kappa\lambda\theta\phi}(x')
\mathcal{P}^{\mu\nu}_{\alpha\beta \gamma\delta}(x) 
\bigg[10 \widehat{\mathcal{F}}_{2}
\mathcal{T}^{\alpha\kappa}\mathcal{T}^{\beta\lambda}
\mathcal{T}^{\gamma\theta}\mathcal{T}^{\delta\phi}
- \frac{4}{H^2} \widehat{\mathcal{F}}'_{2}
\frac{\partial y}{\partial x_{\alpha}}\frac{\partial y}{\partial x'_{\kappa}}
\mathcal{T}^{\beta\lambda}
\mathcal{T}^{\gamma\theta}\mathcal{T}^{\delta\phi} \bigg]
}
\nonumber \\
& & 
=
\frac{\partial y}{\partial x'_{\kappa}}\frac{\partial y}{\partial x'_{\theta}}
\mathcal{T}^{\lambda(\mu}\mathcal{T}^{\nu)\phi} f(y)
\delta C_{\kappa\lambda\theta\phi}(x')
\;.
\qquad \qquad \qquad \qquad \qquad \qquad \qquad
\label{act_outerP_sim}
\end{eqnarray}
Here the function $f(y)$ is,
\begin{eqnarray}
f(y) \!=\!
-50 \widehat{\mathcal{F}}_{2} 
\!+\! 60(2\!-\!y) \widehat{\mathcal{F}}'_{2} 
\!-\!(40-62y +\frac{31}{2}y^2) \widehat{\mathcal{F}}''_{2} 
\!-\!(2\!-\!y)(4y\!-\!y^2) \widehat{\mathcal{F}}'''_{2} \;.
\end{eqnarray}

The final reduction is accomplished by one more partial integration. 
Let us define the integral $I[f]$ of a function $f(y)$ by the relations,
\begin{equation}
\frac{\partial y}{\partial x'_{\kappa}} f(y) \equiv 
\frac{\partial }{\partial x'_{\kappa}} I[f](y) \quad \mbox{such that} \quad 
\frac{\partial I[f] }{\partial y} = f(y) \;.
\end{equation}
Then the one loop source becomes,
 \begin{eqnarray}
\lefteqn{ \int \!\! d^4x' \, \Bigl[\mbox{}^{\mu\nu}
\Sigma^{\rho\sigma}\Bigr](x;x') h^{(0)}_{\rho\sigma}(x')}
\nonumber \\
& & \hspace{-.5cm}
= 2i\sqrt{\!-\!g(x)} 
\int \!\! d^4x' \!\sqrt{\!-\!g(x')}
\frac{\partial y}{\partial x'_{\kappa}}f(y)
\frac{\partial y}{\partial x'_{\theta}}
\mathcal{T}^{\lambda(\mu}\mathcal{T}^{\nu)\phi} 
\delta C_{\kappa\lambda\theta\phi}(x')
\\
& & \hspace{-.5cm}
= - 2i\sqrt{\!-\!g(x)} 
\int \!\! d^4x' \!\sqrt{\!-\!g(x')}I[f]\Bigg\{
\frac{D^2 y}{D x'_{\kappa} D x'_{\theta} }
\mathcal{T}^{\lambda(\mu}\mathcal{T}^{\nu)\phi}\delta C_{\kappa\lambda\theta\phi}(x')
\nonumber \\
& & \hspace{.5cm}
+ \frac{D \mathcal{T}^{\lambda(\mu}\mathcal{T}^{\nu)\phi}}{D x'_{\kappa}}
\frac{\partial y}{\partial x'_{\theta}}\delta C_{\kappa\lambda\theta\phi}(x')
+ \frac{\partial y}{\partial x'_{\theta}}
\mathcal{T}^{\lambda(\mu}\mathcal{T}^{\nu)\phi} 
D^{\kappa}\delta C_{\kappa\lambda\theta\phi}(x')
\Biggr\}
\;.
\label{oneloop_finalred}
\end{eqnarray}
The first and second terms include the metric, 
\begin{equation}
 \frac{D^2 y}{D x'_{\kappa} D x'_{\theta} } = H^2(2-y)g^{\kappa\theta}(x')
,\quad
 \frac{D \mathcal{T}^{\lambda(\mu}\mathcal{T}^{\nu)\phi}}{D x'_{\kappa}}
 = \frac{1}{2}\frac{\partial y}{\partial x_{(\mu}}
\mathcal{T}^{\nu)(\phi}g^{\lambda)\kappa}(x')\;,
\end{equation}
so they give zero when contracted into the linearized Weyl tensor.
The third term vanishes by the transversality of the linearized Weyl tensor
(for dynamical gravitons only) which we showed in (\ref{fullID1}). Hence the
one loop source term for a dynamical graviton is zero:
 \begin{eqnarray}
\int \!\! d^4x' \, \Bigl[\mbox{}^{\mu\nu}
\Sigma^{\rho\sigma}\Bigr](x;x') h^{(0)}_{\rho\sigma}(x') = 0
\;.
\label{oneloop_is_zero}
\end{eqnarray}

Before concluding we should comment on the validity of our result
(\ref{oneloop_is_zero}), in view of the enormous difference between
de Sitter and the actual expansion history of the universe. Of course
equation (\ref{lineqn}) is correct for any geometry, but we only know
the graviton self-energy for de Sitter background. This does not 
make any difference for cosmologically observable tensor perturbations
for two reasons:
\begin{itemize}
\item{As explained section 2.1, de Sitter is an excellent approximation
to primordial inflation up until cosmologically observable perturbations
experience first horizon crossing. After this time the de Sitter
approximation breaks down, but those perturbations are almost constant.}
\item{Our result (\ref{oneloop_step2}) is valid for any geometry, and
the linearized Weyl tensor vanishes for constant perturbations. So there
is no contribtuion from the portion of the integration which derives
from times after the end of inflation.}
\end{itemize}
To see the second point, note that general coordinate invariance 
requires matter contributions to the graviton self-energy to take the 
form (\ref{ansatz}), provided one uses expressions (\ref{Rexp}-\ref{Cexp})
to define the projectors for a general metric, and provided the 
general form of expression (\ref{Tak}) is related to the geodetic
length function through (\ref{length}). That form is all we required
to derive equation (\ref{oneloop_step2}).

\section{Conclusions}

We have found that the inflationary production of MMC scalars has no
effect on dynamical gravitons at one loop order. There is nothing
very surprising about this result. It is exactly what happens in
flat space \cite{HV}. Although the scalar contribution to the
graviton self-energy is enormously more complex in de Sitter than in
flat space, we showed in section 3 that all of this complexity can
be absorbed into surface integrations over the initial time. It is
plausible that these surface integrations can be regarded as 
perturbative redefinitions of the initial state which involve two
scalars and one graviton. The null effect of flat space certainly
has this interpretation, which implies the same for the highest 
derivative part of the de Sitter result. What has yet to be proved
--- and so must be labeled a conjecture --- is that the lower derivative,
intrinsically de Sitter parts have the same interpretation. Checking
this requires a computation like that recently completed for the
self-interacting scalar \cite{KOW1}. 

That is the math behind our result; the physics is that ultraviolet
virtual scalars affect gravitons the same as in flat space, and 
infrared scalars carry too little stress-energy to have much effect.
The effect of ultraviolet scalars is limited, as on flat space, to 
inducing higher derivative counterterms. Although primordial
inflation produces many scalars, they are all highly infrared so
they interact only weakly with gravtions. (This seems to be why 
inflationary gravitons have no significant effect on MMC scalars 
\cite{KW2}.) One might worry that a very infrared graviton would
still suffer some effect from absorbing a comparably infrared 
scalar. To understand why this is not so, let us model the process
by simply replacing the graviton's co-moving wave number $k$ with
a new one $k'$,
\begin{equation}
0 = \ddot{u}(t,k) + 3 H \dot{u}(t,k) + \frac{k^2}{a^2(t)} u(t,k) 
\longrightarrow \ddot{u}(t,k) + 3 H \dot{u}(t,k) + 
\frac{{k'}^2}{a^2(t)} u(t,k) \; .
\end{equation}
The effect on the mode function is negligible after both $1/a^2$
terms have redshifted into insignificance.

Both math and physics suggest that inflationary gravitons might do
something interesting to other gravitons. The graviton contribution
to the graviton self-energy has been derived at one loop order
\cite{TW1} so the computation can be made. Of course one can reduce
the effect to a temporal surface term, as we did in section 3, but
it seems likely that this surface term will depend upon the
observation time $\eta$ so that it cannot be absorbed into a
perturbative correction to the initial state. The reason for this is
that the graviton contribution contains de Sitter-breaking, infrared
logarithms \cite{TW1}, unlike the scalar contribution. The physical
principle involved would be that gravitons possess spin and even
very infrared gravitons continue to interact via the spin-spin
coupling which doesn't exist for scalars. This is presumably why
inflationary gravitons induce a secular enhancement of the field
strength of massless fermions \cite{MW}.

It would also be interesting to investigate how inflationary scalars
affect the force of gravity. That can be done by solving
(\ref{oneloop}) to correct for the linearized response to a
stationary point mass $M$ \cite{TW4},
\begin{equation}
h^{(0)}_{00}(x) = a^2 \times \frac{2 G M}{a \Vert \vec{x}\Vert} \; ,
\; h^{(0)}_{0i}(x) = 0 \; , \; h^{(0)}_{ij}(x) = a^2 \times \frac{2
G M}{a \Vert \vec{x} \Vert} \times \delta_{ij} \; . \label{dSS}
\end{equation}
The same reduction procedures we laid out in section 3 can be
applied in this case except that:
\begin{itemize}
\item{The spin zero projector $\mathcal{P}^{\rho\sigma}(x')$ does not
annihilate (\ref{dSS}); and}
\item{The linearized stress tensor does not vanish.}
\end{itemize}
Because the linearized stress tensor is proportional to
$\delta^3(\vec{x'})$, we should be able to reduce the computation to
a single integration over $\eta'$.

Note that the virtual scalars of flat space do induce a correction
to the classical potential \cite{EMGR,us} and we expect one as well
on de Sitter background. On dimensional grounds the flat space
result must (and does) take the form,
\begin{equation}
\Phi_{\rm flat} = -\frac{GM}{r} \Biggl\{1 + {\rm constant} \times
\frac{G}{r^2} + O(G^2) \Biggr\} \; .
\end{equation}
On de Sitter background there is a dimensionally consistent
alternative provided by the Hubble constant $H$ and by the secular
growth driven by continuous particle production,
\begin{equation}
\Phi_{\rm dS} = -\frac{GM}{r} \Biggl\{1 + {\rm constant} \times  G
H^2 \ln(a) + O(G^2) \Biggr\} \; .
\end{equation}
If such a correction were to occur its natural interpretation would
be as a time dependent renormalization of the Newton constant. The
physical origin of the effect (if it is present) would be that
virtual infrared quanta which emerge near the source tend to
collapse to it, leading to a progressive increase in the source.

\vskip .5cm

\centerline{\bf Acknowledgements}

This work was partially supported by NSF grant PHY-0855021 and by
the Institute for Fundamental Theory at the University of Florida.

\end{document}